# Development and Validation of a Faculty Artificial Intelligence Literacy and Competency (FALCON-AI) Scale for Higher Education

Yukyeong Song [a] [*]
Assistant Professor
E-mail: ysong51@utk.edu

Hyunjoo Moon [b]
Ph.D. Candidate
E-mail: hm342821@ohio.edu

Hyewon Yang [c]
Research Professor
Email: hwyang@ssu.ac.kr

Chris Kilgore [d]
Director
Email: ckilgor4@utk.edu

[a] Department of Theory and Practice in Teacher Education, University of Tennessee, Knoxville, TN 37916, USA

[b] Patton College of Education, Ohio University, Athens, OH 45701, USA

[c] Center for Teaching and Learning Innovation, Soongsil University, Seoul, 06978 South Korea

[d] Teaching and Learning Innovation, University of Tennessee, Knoxville, TN 37996, USA

*Corresponding Author

# Abstract

The integration of artificial intelligence (AI) in higher education underscores the growing importance of faculty AI literacy and competency across teaching, research, and service. Existing AI literacy instruments, however, primarily target the general public, students, or K-12 teachers, and therefore lack the role-embedded indicators and psychometric validation needed for scalable assessment among university faculty. Grounded in the Critical Tech-resilient Literacies (CTRL) framework, this study develops and validates the Faculty Artificial Intelligence Literacy and Competency (FALCON-AI) Scale as a concise and practically deployable tool for higher education contexts. Using a theory-driven development process, we generated an initial pool of 43 items mapped to three literacies (functional, evaluative, and ethical literacy) and situated them across four faculty work domains (general, teaching, research, service/administration), creating a 3 × 4 framework. Content validation was conducted through structured interviews with four subject-matter experts, supplemented by a GPT-based reviewer to triangulate ratings of clarity, relevance, and necessity, yielding refined 39 items for pilot testing. Pilot testing involved 269 valid responses, which were analyzed using confirmatory factor analysis (CFA). CFA evaluated the theoretically specified structure, followed by item reduction to minimize respondent burden while preserving content coverage. The final 23-item FALCON-AI demonstrated good model fit for the AI Literacy x Faculty Work measurement and strong reliability. This study presents a validated FALCON-AI scale with good reliability and validity, offering a refined practical instrument for assessing faculty AI in higher education.

**Keywords:** AI literacy, University faculty, Measurement, Scale, Higher education

## 1. Introduction

Artificial Intelligence (AI) is fundamentally reshaping the landscape of higher education, transforming societal structures, academic research, teaching and learning, and the professional responsibilities of faculty (Adamakis & Rachiotis, 2025). In the instructional domain, AI is being integrated to personalize learning pathways, automate grading, enhance instructional efficiency, and elevate students' digital competencies (Ning et al., 2025). Simultaneously, AI is revolutionizing research methodologies, facilitating tasks ranging from automated literature mapping and data analysis to academic writing and editing (Zhou & Schofield, 2024). Beyond these core academic functions, AI offers significant potential to streamline administrative workflows, allowing faculty to automate repetitive tasks and optimize productivity (Adamakis & Rachiotis, 2025).

To adapt to this rapid transition, institutions are increasingly formalizing AI policies for course management and assessment, and academic journal publishers are exploring ways to maintain academic integrity, with some of them now mandating the disclosure of AI use in scholarly work (Adamakis & Rachiotis, 2025; Ganjavi et al., 2024). However, compliance remains a challenge, leaving the higher education society largely dependent on voluntary engagement, ethical awareness, and AI literacy levels of its participants (Liang et al., 2025). In this context, faculty members play a critical role in the responsible, ethical, and discerning use of AI, demonstrating professionalism to enhance productivity, improve students' learning, and support their research.

In this background, it is important to foster AI literacy among faculty in higher education. AI literacy is defined as the ability to understand, use, and critically evaluate AI in societal and educational contexts (Long & Magerko, 2020). For higher education faculty, this literacy is multidimensional, spanning the teaching, research, and service sides of faculty jobs (Kennedy & Gupta, 2025). Crucially, AI literacy emphasizes technology as a collaborator or an augmentation of human intelligence rather than a replacement for professional roles (Tzirides et al., 2024). By fostering this literacy, faculty can enhance their productivity and focus on high-value cognitive tasks and meaningful interactions with students, ensuring that human oversight remains central to academic and instructional work (Adamakis & Rachiotis, 2025; Tzirides et al., 2024).

Given the critical importance of these competencies, researchers have developed AI literacy assessment tools in recent years. Existing scales are designed for various target populations, such as the general public (e.g., Wang et al., 2023), K-12 learners (e.g., Su, 2024), K-12 teachers (e.g., Ning et al., 2025), or higher education students (e.g., Biagini et al., 2025; Hwang et al., 2023). While those scales include essential competency statements for their target audience within generalizable constructs of AI literacy, there is no existing validated, ready-to-use scale contextualized for university faculty. Validated measurement tools will allow faculty to assess their current status, set professional goals, and identify necessary professional development (PD) opportunities for AI literacy development (Ning et al., 2025; Zary, 2024). Furthermore, as institutions implement workshops to foster AI literacy among faculty and staff, pre- and post-assessments are vital for evaluating the educational impact of these interventions and supporting long-term data-informed decision-making (Ning et al., 2025).

To this purpose, this study aims to develop and validate the Faculty Artificial Intelligence Literacy and Competency (FALCON-AI) Scale for Higher Education, a concise instrument designed to measure the practical application of AI across teaching, research, and administrative domains.

## 2. Literature review

### 2.1. AI literacy frameworks

AI literacy is defined as the ability to understand, use, monitor, and critically evaluate AI applications in social and educational context, encompassing not only technical skills but also ethical reasoning (Laupichler et al., 2022; Long & Magerko, 2020). AI literacy frameworks have been developed across various educational levels with distinct areas of focus depending on the target audience. In the K-12 context, the AI4K12 Five Big Ideas provide foundational content of AI to teach to K-12 students (i.e., Perception, Representation & Reasoning, Learning, Natural interaction, Societal impact) (Touretzky et al., 2019). For K-12 educators, models often adapt the Technological Pedagogical Content Knowledge (TPACK) framework into variations like AI-TPACK or AI-PACK, which emphasize the intersection of AI knowledge with pedagogical skills and subject matter expertise (Dogan et al., 2025; Lorenz & Romeike, 2023). These K-12 frameworks provide general insights into faculty AI literacy by highlighting that teaching competency requires not just technical skill, but the ability to integrate AI tools into instructional design and pedagogical practice (Ning et al., 2025).

Meanwhile, AI literacy has been discussed in the higher education context as well. Frameworks developed for higher education have largely prioritized student competencies to ensure workforce readiness and academic integrity. Hackl et al. (2026) proposed the "AI Literacy Heptagon" framework to guide the development of college students' competencies and the evaluation of higher education curricula. The framework includes seven distinct dimensions: Technical Knowledge and Skills, Application Proficiency, Critical Thinking Ability, Ethical Awareness and Reasoning, Social Impact Understanding, Integration Skills, and Legal and Regulatory Knowledge. Similarly, Kennedy and Gupta (2025) introduce the "AI & Data Acumen Learning Outcomes Framework," designed to scaffold holistic literacy by intersecting cognitive processes with specific knowledge domains. This model is structured as a matrix combining four proficiency levels (from Foundational to Expert) with seven knowledge dimensions: Self-efficacy, Ethics, Collaboration, Socio-cultural, Innovation and Creativity, Cognitive, and Technical. Furthermore, Zhou and Schofield (2024) propose a conceptual framework designed to support module organizers and educators in integrating AI into curricula using a bottom-up approach, adopting Ng et al. (2021)'s suggested components of AI literacy: Know and understand AI, Use and apply AI, Evaluate and create AI, and AI ethics. Despite these active discussions in higher education, existing frameworks focus on college students as the target population or on instructors' teaching competencies.

On the other hand, when developing AI literacy measurement scales, previous literature has adopted more general constructs to operationalize these competencies. For example, Ning et al. (2025) categorized the AI literacy constructs into Perception, Knowledge and skills, Application and innovation, and Ethics when developing the Artificial Intelligence Literacy Scale for Teachers (AILST). Wang et al. (2023) adopted Awareness, Usage, Evaluation, and Ethics when they developed and validated the AI literacy scale (AILS) for the general public. Similarly, Biagini et al. (2025) grounded their Critical Artificial Intelligence Literacy Scale (CAILS) for higher education students in four consolidated dimensions: Knowledge-related, Operational, Critical, and Ethical. Drawing upon these previous studies, we adopted the Critical Technology Resilient Literacies (CTRL) framework as a general framework to represent the essential competencies of AI literacy specifically for university faculty.

## 2.2. The Critical Tech-Resilient Literacies (CTRL) Framework

This study adopts the Critical Tech-Resilient Literacies (CTRL) framework to specifically address the professional needs of higher education faculty. CTRL is a multiliteracy framework designed to support instructors in using and evaluating generative AI software, as well as in designing learning activities that deploy these tools effectively. The framework builds upon the theoretical foundation laid by Gupta et al. (2024), which was itself adapted from Selber's (2004) earlier multiliteracies model for the digital age. Incorporating a critical media literacies lens (Kellner & Share, 2005), CTRL shares the three-tiered structure of these antecedents but reorients them into a "tech-resilient" praxis designed to withstand the rapid obsolescence of specific software tools.

The framework identifies three distinct but interrelated literacies required when engaging with AI: functional literacy, evaluative literacy, and ethical literacy.

*Functional literacy* refers to the user's ability to understand and use software to achieve optimal performance for a specific purpose in a given context. Functional literacy includes the ability to answer questions such as, "How does this system work?" and "How can I get the most out of this system?" *Evaluative literacy* refers to the user's ability to discern the strengths and weaknesses of specific software and determine the optimal software for a given purpose. This dimension emphasizes what Gupta et al. (2024) envisioned as an "informed critique," where a user needs to test and critically evaluate the value of a given software. Evaluative literacy includes the ability to answer questions such as, "What are this system's strengths and weaknesses (for this purpose, in this context)?" "How do I (come to) trust this system?" and "What biases or other hidden influences does this technology embody based on how it was created?" Finally, *ethical literacy* identifies the user's ability to make informed decisions about *whether* and *how* to use a given system. Generative AI technology presents complex ethical challenges; using a system might provide immediate benefits with little apparent harm, yet the downstream aggregate consequences of widespread use may outweigh those benefits. Ethical literacy involves addressing questions such as, *should I use this system (for this purpose, in this context)?, what would happen if everybody used this system (for similar purposes)?,* and *who would benefit or be harmed if people used this system?*

We adopted the CTRL framework because it is comprehensive yet concise, prioritizing the essential competencies required for faculty work. While many existing frameworks separate theoretical understanding from application (e.g., Ng et al., 2021), the CTRL framework's first component, *Functional Literacy*, encompasses both practical knowledge and operational application. This integration better represents AI competencies required for faculty, for whom the practical ability to navigate and utilize AI tools often takes precedence over theoretical understanding. Furthermore, while previous frameworks envisioned separate, parallel literacies or lenses (Gupta et al., 2024), the CTRL framework understands these three levels as scaffolded layers, "stacked" on top of each other for analytical purposes. For example, to evaluate a piece of software (evaluative literacy), a user must first possess the functional literacy to use it; to make informed ethical decisions (ethical literacy), a user must be able to address key questions at both the functional and evaluative levels.

As illustrated in Table 1, the three layers of the CTRL framework align closely with, and synthesize, the diverse constructs found in existing AI literacy models (e.g., Hackl et al., 2026; Long & Magerko, 2020; Ng et al., 2021), providing a unified lens for assessing faculty competency.

**Table 1**

*The Adopted CTRL Framework: Constructs, Definitions, and Alignment with Existing Frameworks*

| Literacy | Ability | Grounding AI Literacy Frameworks |
|---|---|---|
| Functional Literacy | Ability to understand and use AI technologies | What is AI & How does AI work? (Long & Magerko, 2020) Know and understand, & Use and apply (Ng et al., 2021); AI awareness & AI use (Wang et al., 2023) Technical knowledge and skills, Application proficiency & Integration Skills (Hackl et al., 2026); Cognitive & Technical (Kennedy & Gupta, 2025); Knowledge-related & Operational (Biagini et al., 2025) |
| Evaluative Literacy | Ability to evaluate AI technologies' strengths and weaknesses and determine appropriate applications of AI technologies for context-specific needs | What can I do? (Long & Magerko, 2020); Evaluate AI (Ng et al., 2021); AI evaluation (Wang et al., 2023); Critical Thinking Ability, Legal and Regulatory Knowledge (Hackl et al., 2026); Critical (Biagini et al., 2025) |
| Ethical Literacy | Ability to make an informed decision about where, when, and how to use AI technology, complying with ethical guidelines | How should AI be used? (Long & Magerko, 2020); Ethical issues (Ng et al., 2021); AI ethics (Wang et al., 2023); Ethical Awareness and Reasoning & Social Impact Understanding (Hackl et al., 2026); Ethics (Kennedy & Gupta, 2025); Ethical (Biagini et al., 2025) |

## 2.3. AI literacy assessment

Recent scholarship has produced multiple instruments to measure AI literacy across diverse populations, reflecting growing recognition of AI as a critical competency in contemporary society. Celik (2023) developed a 27-item scale to measure K-12 teachers' knowledge of AI-based instruction. The instrument comprises five factors: Intelligent Technological Knowledge (ITK), Intelligent Technological Pedagogical Knowledge (ITPK), Intelligent Technological Content Knowledge (ITCK), Intelligent Technological Pedagogical Content Knowledge (ITPACK), and Ethics. These factors are grounded in an extended TPACK framework that integrates ethical considerations into the traditional TPACK model, thereby emphasizing responsible and reflective AI integration in teaching practice. While this instrument makes an important contribution to understanding AI-related pedagogical knowledge among school teachers, its conceptualization remains situated within K-12 instructional contexts.

Wang et al. (2023) proposed the Artificial Intelligence Literacy Scale (AILS), conceptually grounded in the framework of digital literacy. The instrument was designed to assess general AI user competence among ordinary users rather than domain-specific professionals. It is recommended for measuring overall AI literacy instead of proficiency in using particular AI applications. The AILS consists of 12 items organized into four constructs: awareness, use, evaluation, and ethics. Similarly, Laupichler et al. (2023) developed the Scale for the Assessment of Non-Experts' AI Literacy (SNAIL) to measure AI literacy among general individuals without specialized technical backgrounds. The authors identified three underlying competency dimensions: Technical Understanding, Critical Appraisal, and Practical Application. The 31-item instrument was intentionally designed to focus exclusively on AI literacy, distinguishing it from related constructs such as digital literacy, an approach that contrasts with digital literacy-grounded models (e.g., Wang et al., 2023).

Expanding the scope to adult populations more broadly, Soto-Sanfiel et al. (2025) developed the 56-item Scale of Artificial Intelligence Literacy for All (SAIL4ALL) as a comprehensive instrument to assess AI literacy across adult populations. Grounded in the conceptual framework proposed by Long and Magerko (2020), the scale is organized around four core domains: understanding what AI is, what AI can do, how AI works, and how AI should be used. Unlike many existing AI literacy instruments that rely on self-reported perceptions of skills, experiences, or attitudes, SAIL4ALL focuses on measuring factual knowledge. The instrument employs a two-point true/false response format to evaluate respondents' understanding of factual statements about AI. Designed for use with the general population, SAIL4ALL provides a knowledge-based assessment approach that minimizes subjective bias and offers a more direct measure of conceptual AI understanding.

Pinski and Benlian (2023) proposed a general AI literacy scale aimed at measuring human competency in artificial intelligence. Their instrument comprises 13 items across five dimensions: AI technology knowledge, knowledge of human actors in AI, knowledge of AI processes and steps, AI usage experience, and AI design experience. The scale was developed to assess the overall level of general AI literacy, with particular emphasis on individuals' perceived competencies. Since the instrument targets general AI literacy rather than role-specific responsibilities, it lacks sensitivity to the contextual and accountability-driven competencies required in professional domains.

Ning et al. (2025) developed the Artificial Intelligence Literacy Scale for Teachers (AILST) to assess AI literacy among both K-12 and higher education educators. The final instrument consists of 36 items organized into four dimensions: AI perception, AI knowledge and skills, AI applications and innovation, and AI ethics. While the AILST provides a comprehensive framework for evaluating teachers' readiness to understand and integrate AI into instructional practice, it does not fully capture the broader and more specialized responsibilities of faculty members in higher education.

More recently, Chen et al. (2025) developed a co-constructed generative AI Literacy survey as part of the AI Academy, a faculty development program designed for higher education instructors. The instrument targeted university faculty and was used as a pre-post measure to examine changes in self-assessed AI literacy. Following participatory revision, the final survey consisted of 36 items organized into subscales capturing competencies such as knowledge of generative AI models, understanding of AI capacity and limitations, skills in using generative AI tools for teaching, the ability to detect and assess AI outputs, prompt engineering skills, contextual knowledge, ethical implications, legal aspects, and continual learning. Although the instrument demonstrates a participatory design for practical use, the developed scale does not comprehensively encompass faculty work as a whole, but rather focuses more on teaching competency. Moreover, the developed scale lacks full psychometric validation, which limits its construct validity, generalizability, and utility as a standardized measure of faculty AI literacy.

Taken together, these instruments demonstrate important conceptual advancements in defining and operationalizing AI literacy. However, existing scales predominantly target the general population, non-experts, or K-12 teachers and few instruments are specifically designed to capture the multifaceted and role-embedded responsibilities of higher education faculty, including accountability in research, pedagogical decision-making in AI-mediated instructional design, assessment redesign in the age of generative AI, and ethical leadership in guiding students' responsible AI use. Moreover, some faculty-focused efforts have prioritized developmental reflection over validation, limiting their applicability as generalizable assessment tools. This gap underscores the need for a validated, faculty-focused AI literacy scale that reflects the unique professional responsibilities of university instructors across teaching, research, and service domains. In response to this need, our study aims to develop a contextually grounded and validated measure of faculty AI literacy.

## 3. Methods

This study aims to develop a valid and reliable measure for higher education faculty's AI literacy and competency (FALCON-AI). FALCON-AI was developed following the procedure, including initial item generation, expert content validation, pilot testing, confirmatory factor analysis (CFA), and reliability testing. This study was reviewed and approved by the Institutional Review Board at the first author's university (Anonymized for review).

### 3.1. Initial scale development

We adopted the CTRL framework to define AI literacy for higher education faculty with three main constructs: functional literacy, evaluative literacy, and ethical literacy (Authors et al., underreview, see section 2.2. for more details). In addition, we categorized higher education faculty's work into three areas: teaching, research, and service/administration, following widely adopted conventions in academia (e.g., Moore & Ward, 2010). There are also general tasks that do not fall into specific faculty work categories. Therefore, we created a framework that consists of 12 dimensions (3 literacy * 4 faculty work) (see Table 2).

**Table 2**

*FALCON-AI scale development framework*

| AI literacy \ Faculty work | General | Teaching | Research | Service |
|---|---|---|---|---|
| Functional Literacy | | | | |
| Evaluative Literacy | | | | |
| Ethical Literacy | | | | |

Based on this 3 x 4 framework, we reviewed previous literature on the AI literacy scales, such as Celik (2023), Wang et al. (2023), Pinkski & Benlian (2023), Ning et al. (2025), Chen et al. (2025) (see section 2.3 for more details), and filled the table (Table 2) with applicable existing items. While some items were retained from the original statement, others were revised to be more contextualized in higher education. For example, the item "Based on instructional content, I can select appropriate teaching methods and AI tools for instruction," retrieved from Ning et al (2025) was revised to "I can compare different AI tools and determine which ones to use for different teaching tasks (e.g., lesson design, student feedback)." We also created new items if there are no applicable items from the existing scales (e.g., "I can include a clear AI guideline in my syllabus detailing how students may or may not use AI for assignments and assessments"). Through this process of borrowing, revising, and creating, we included 43 items in the initial item list.

## 3.2. Expert content validation

To assess the clarity, relevance, and necessity of the initial 43 items, we conducted a content validation study with subject matter experts (SMEs). Four experts from related fields participated in the content validation of the initial scale (see Table 3).

**Table 3**

*Subject Matter Experts Participating in Content Validation*

| Experts | Title & Institution | Field of expertise | Years of experience Post-Ph.D. |
|---|---|---|---|
| A | Ph.D., Professor, R1 University | Universal Design for Learning, AI Education, Computer Science Education | 18 years |
| B | Ph.D., Assistant Professor, R1 University | AI in Education, Human-AI interaction | 4 years |
| C | Ph.D., Assistant Professor, R1 University | AI engineering, Computer Science | 3 years |
| D | Ph.D., Director of R1 University's Center for Teaching and Learning | Faculty professional development related to AI | 16 years |

Researchers conducted one-on-one interviews with the four experts via Zoom. To elicit experts' immediate and unbiased reactions, mirroring the first-time exposure that target users would have to the scale, and to minimize the use of AI tools during feedback generation, the scale items were not shared in advance. Instead, items were presented via real-time screen sharing during the interviews. All meetings were audio- and video-recorded and subsequently transcribed to capture detailed expert feedback. During the interviews, researchers also recorded experts' validity ratings directly on the shared document.

We adopted the approach proposed by Zamanzadeh et al. (2015), in which experts review and rate each item systematically. Each item was evaluated across three dimensions (i.e., clarity, relevance, and necessity) using a 3-point rating scale (see Table 4). In addition to providing numerical ratings, experts were asked to offer qualitative feedback and concrete suggestions for item improvement. Specifically, for items rated low in clarity (i.e., score = 1), experts were asked to recommend revisions to improve wording and better convey the intended meaning. For items receiving low relevance ratings, we discussed whether the items should be reassigned to different constructs or removed from the scale. For items rated low in necessity, we asked experts to give suggestions on revising the items, integrating them with other related items, or eliminating them.

Although many prior studies have employed five SMEs for content validation (e.g., Wang et al., 2023), only four experts participated in this study. To complement the human expert review, we additionally employed an AI-based evaluation to provide supplementary validation scores and qualitative feedback for each item. Specifically, we used OpenAI's ChatGPT 5.2 Thinking model (OpenAI, 2026). The model was provided with the same content validation document used in the human SME interviews, along with the following prompt:

> **Prompt:**
> *You are an expert in AI in Education with experience as a faculty member who uses AI in teaching, research, service, and administration. You are asked to conduct a content validation of a newly developed AI literacy scale for faculty. Please provide feedback on the clarity, relevance, and necessity of each item, and suggest how items might be revised. Feedback may include rewording for clarity, moving an item to a different construct (e.g., from evaluative literacy to ethical literacy), removing an item, splitting an item into multiple items, or proposing new items.*

Based on the content validation from the human SMEs and AI supplement, two authors reviewed the item ratings, discussed discrepancies among expert evaluations, and considered potential revisions. Final decisions were then made regarding the scale items to be included in the pilot study. As a result, 39 items were finalized for pilot testing.

**Table 4**

*Content Validation Criteria and Rating Scales*

| Criterion | Definition | Scores |
|---|---|---|
| Clarity | The extent to which the description in each item is clear to the target audience (university faculty) | 1 (not clear), 2 (item needs minor revision to be clear), 3 (very clear) |
| Relevance | the extent to which each item is related to the construct | 1 (not relevant), 2 (item needs minor revision to be relevant), 3 (very relevant) |
| Necessity | The extent to which each item is | 1 (not necessary), |

| | necessary to measure the construct | 2 (useful but not essential), |
| | | 3 (essential) |

The content validity index (CVI) was used to evaluate item relevance and clarity. Experts rated each item on a three-point scale (1 = not relevant/not clear, 2 = needs minor revision, 3 = highly relevant/very clear). For CVI calculation, ratings of 2 or 3 were considered acceptable. Item-level content validity indices (I-CVIs) were computed by dividing the number of experts assigning a rating of 2 or 3 by the total number of experts. Scale-level indices (S-CVI) were also calculated to assess overall instrument validity.

To assess the necessity of each item, the content validity ratio (CVR) was calculated following Lawshe's (1975) method. Experts were asked to indicate whether each item was "not necessary (score = 1)," "useful but not essential (= 2)," or "essential (= 3)" for representing the intended construct. CVR values range from –1 to 1, with higher values indicating stronger expert agreement regarding item essentiality. Because the original Lawshe critical values have been shown to be overly conservative for small expert panels, the revised critical values proposed by Ayre and Scally (2014) were adopted. For a panel of five experts, the minimum acceptable CVR value is 0.49 at the 0.05 significance level. Accordingly, items with CVR values equal to or greater than 0.49 were retained, whereas items falling below this threshold were considered for removal or substantive revision.

### 3.3. Pilot testing

We conducted a pilot test to collect responses to the expert-validated scale for factor analyses. The survey was administered via Qualtrics, allowing the electronic instrument to be easily distributed and completed on participants' computers or smartphones. FALCON-AI was formatted in the form of a five-point Likert scale (Likert, 1932). Participants were recruited through social media, institutional listservs, and the researchers' professional networks, targeting university faculty, graduate assistants, postdoctoral researchers, and university staff. Although FALCON-AI primarily targets university faculty, the underlying construct of interest is AI literacy in higher education across teaching, research, and service and administration. Accordingly, we included graduate assistants and postdoctoral researchers, who engage in similar activities as part of their professional training, as well as university staff who support faculty work in these domains.

Of the 288 participants who began the online survey, 19 responses were excluded due to dropout or incomplete data, resulting in a final sample of 269 participants. Regarding educational attainment, the majority of participants held a doctoral degree (75.8%, n = 204), followed by a master's degree (19.7%, n = 53), and a bachelor's degree (4.5%, n = 12). Participants also represented a range of employment statuses. The largest group consisted of tenure-track faculty

(26.8%, n = 72), followed by teaching faculty (22%, n = 59) and graduate students serving as teaching assistants or research assistants (20.8%, n = 56). In terms of years of experience, 35.32% of participants reported having 0–5 years of experience, followed by 6–10 years (27.88%, n = 75) and more than 20 years (20.45%, n = 55). Most participants were affiliated with research-intensive universities (63.57%, n = 171), followed by teaching-focused universities (33.83%, n = 91), with very few participants from community colleges and other institutional types.

**Table 5**

*Participant information in the pilot study*

| Variables | Catetoreis | Number | Proportion |
|---|---|---|---|
| Highest Education | Bachelor's Degree | 12 | 4.5% |
| | Master's Degree | 53 | 19.7% |
| | Doctoral Degree | 204 | 75.8% |
| Employment Status | Tenure-track faculty | 72 | 26.8% |
| | Tenured faculty | 35 | 13.0% |
| | Teaching faculty | 59 | 22% |
| | Research faculty | 9 | 3.3% |
| | Part-time faculty | 15 | 5.6% |
| | Post-doctoral researcher | 11 | 4.1% |
| | Graduate students | 56 | 20.8% |
| | University Staff | 12 | 4.5% |
| Discipline | Education | 151 | 47.63% |
| | Humanities | 54 | 17.03% |
| | Engineering & Technology | 29 | 9.15% |
| | Social Sciences | 28 | 8.83% |
| | Health / Medical Sciences | 16 | 5.05% |
| | Natural Sciences | 13 | 4.10% |
| | Business / Law | 8 | 2.52% |
| | Arts & Design | 4 | 1.26% |
| | Communication & Media | 3 | 0.95% |
| | Agriculture & Environmental Studies | 3 | 0.95% |
| | Other | 8 | 2.52% |
| Years of Experience | 0-5 | 95 | 35.32% |
| | 6-10 | 75 | 27.88% |
| | 11-15 | 27 | 10.04% |
| | 16-20 | 17 | 6.32% |
| | More than 20 | 55 | 20.45% |
| Institutional Context | Research-intensive university | 171 | 63.57% |
| | Teaching-focused university | 91 | 33.83% |
| | Community college | 2 | 0.74% |
| | Other | 5 | 1.86% |

### 3.4. Confirmatory factor analysis

Confirmatory factor analysis (CFA) was conducted to examine the adequacy of the hypothesized measurement structure underlying the FALCON-AI scale. Because this study sought to test a theoretically specified 3 × 4 framework derived a priori from established theoretical foundations, the analysis was confirmatory rather than exploratory in nature (Hinkin, 1998). The proposed structure was grounded in the Critical Tech-Resilient Literacies (CTRL) framework, which conceptualizes AI literacy as comprising three interrelated literacies (i.e., functional, evaluative, and ethical) and was systematically crossed with four domains of faculty work: general, teaching, research, and service/administration. This multidimensional architecture was explicitly defined prior to data collection and operationalized during item development, where each item was intentionally mapped onto a specific literacy-by-role dimension. Therefore, the purpose of the analysis was not to uncover latent factors inductively, as in exploratory factor analysis, but to evaluate whether the observed covariance structure was consistent with the theoretically articulated model.

To establish construct validity, two separate CFA models were estimated: (a) a three-factor model representing the literacy dimension (i.e., functional, evaluative, and ethical literacy), and (b) a four-factor model representing faculty work domains (i.e., general, teaching, research, and service/administration). This validation strategy aimed to verify that each core dimension of the proposed 3 × 4 framework formed a psychometrically sound latent structure. Establishing the validity of these two foundational dimensions provides essential preliminary support for the conceptual integrity of the integrated crossed framework.

Considering the ordinal nature of the data collected using a five-point Likert scale and the potential for non-normality, CFA was performed using the robust maximum likelihood estimator (MLR), which provides robust standard errors and model fit indices.

Model fit was evaluated using multiple indices rather than relying on a single criterion, including $\chi^2$, the Comparative Fit Index (CFI), the Tucker-Lewis Index (TLI), the Root Mean Square Error of Approximation (RMSEA), and the Standardized Root Mean Square Residual (SRMR), in accordance with general recommendations (Kline, 2016). Given the sensitivity of $\chi^2$ to sample size and model complexity, it was interpreted as a reference indicator. There is no absolute cutoff value for model fit indices; rather, commonly cited thresholds serve as general guidelines. Incremental fit indices (CFI and TLI) of .90 or higher were considered indicative of acceptable fit. For absolute fit indices, RMSEA and SRMR values of .05 or below were interpreted as excellent fit, and values up to .08 as acceptable.

### 3.5. Reliability

Internal consistency reliability was assessed using Cronbach's alpha coefficients. Consistent with conventional psychometric standards, alpha values of .70 or higher were considered acceptable for research purposes (Nunnally & Bernstein, 1994). In practice, values above .80 are generally interpreted as good, and values of .90 or higher as excellent. Reliability estimates were examined alongside standardized factor loadings to ensure that each subscale demonstrated adequate internal coherence without evidence of redundancy. All statistical analyses were conducted using R (version 4.5.1).

## 4. Results & Discussion

### 4.1. Content validity

A total of 43 initial items went through content validation by four human experts and one AI agent (i.e., OpenAI's ChatGPT). Content validity indices were computed from expert ratings of this initial item pool. Specifically, item-level Content Validity Indices (I-CVIs) for clarity and relevance were calculated by treating ratings of 2 or 3 on a three-point scale as acceptable. Based on these I-CVI results, 13 items were eliminated due to insufficient expert agreement (I-CVI < .70). In addition, qualitative expert feedback indicated that several items contained multiple conceptual components within a single statement. To improve conceptual precision and measurement clarity, these items were refined and, where appropriate, separated into more specific sub-items.

Following item elimination and item splitting, the instrument was finalized with 39 items. The overall content validity evidence derived from the initial expert review indicated strong agreement among raters, with scale-level CVI values of .94 for clarity and .95 for relevance. Item necessity was evaluated using Lawshe's (1975) Content Validity Ratio (CVR), and the mean CVR across experts was .58, exceeding the minimum acceptable threshold of .49 for a five-expert panel (Ayre & Scally, 2014). Inter-rater agreement (IRA), calculated as the proportion of ratings matching the modal rating across items, ranged from .77 to .85 among only human experts and from .79 to .87 when GPT ratings were included. The resulting 39-item version of the instrument was subsequently advanced to the pilot testing phase for further psychometric evaluation, including CFA and reliability testing.

**Table 6**

*Content Validity*

| Criteria | A | B | C | D | GPT | CVI/CVR | IRA (Human only) | IRA (Human +GPT) |
|---|---|---|---|---|---|---|---|---|
| Clarity | 2.74 | 2.83 | 2.81 | 2.26 | 2.91 | 0.94 | 0.77 | 0.79 |
| Relevance | 2.95 | 2.76 | 2.9 | 2.55 | 2.88 | 0.95 | 0.85 | 0.87 |
| Necessity | 2.86 | 2.76 | 2.81 | 2.52 | 2.65 | 0.58 | 0.85 | 0.81 |

### 4.2. Confirmatory factor analysis

A confirmatory factor analysis (CFA) was conducted on the 39 items that satisfied content validity. Both constructs were modeled as correlated first-order factor models. For AI Literacy, the first three-order factors (i.e., functional, evaluative, and ethical literacy) were specified, with correlations freely estimated among them. For Faculty Work, four first-order factors (i.e., general, teaching, research, and service/administration) were specified, also with freely estimated correlations. This approach was chosen because the constructs share common items and are conceptually related rather than strictly hierarchical. Allowing correlations among first-order factors acknowledges these theoretical relationships while maintaining the interpretability of individual subscales and ensuring consistency in evaluating factor structure and psychometric properties across constructs.

The results of the analysis indicate that for AI Literacy, the chi-square test was $\chi^2$ = 1534.430 (df = 699, p < .001), CFI = 0.836, TLI = 0.826, RMSEA = 0.074 [90% CI: 0.069–0.079], and SRMR = 0.069. Although the CFI and TLI are slightly below the conventional cutoff of 0.90, the RMSEA and SRMR values indicate an acceptable level of approximation, suggesting that the hypothesized factor structure aligns reasonably well with the observed data, but minor refinements could further improve model fit. For Faculty Work, the chi-square test was $\chi^2$ = 1854.696 (df = 696, p < .001), CFI = 0.773, TLI = 0.758, RMSEA = 0.087 [90% CI: 0.082–0.092], and SRMR = 0.079. Compared with AI Literacy, the lower CFI and TLI and higher RMSEA suggest that the model fit is relatively poorer and requires improvement, indicating that the factor structure does not fully capture the observed data patterns and that further refinement of items or factors is necessary.

The final number of items was planned to be around 20 to reduce respondent fatigue. Items with low factor loadings were prioritized for deletion, and the remaining items were reviewed to ensure content relevance and to maintain a balanced number of items between the AI Literacy and Faculty Work constructs. In total, 23 items were selected as the final list of questions. This process aimed to optimize both psychometric properties and practical feasibility of the instrument.

The final measurement model for AI Literacy is presented in Figure 1.

**Figure 1**

*Measurement Model of AI Literacy*

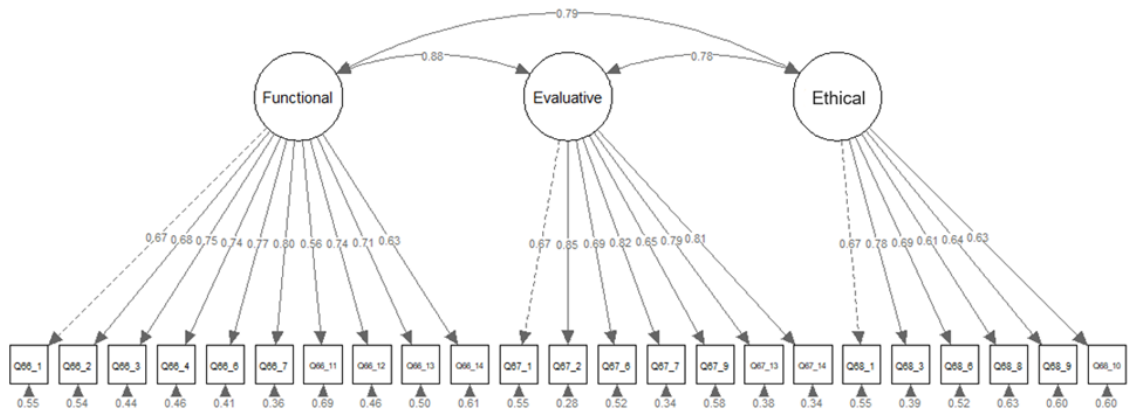

The model fit indices for the AI Literacy measurement model are shown in Table 7. Although the chi-square test was statistically significant, the CFI and TLI were both above 0.90, and the RMSEA and SRMR were below 0.08. Taken together, these indices indicate that the hypothesized model demonstrates a good fit to the observed data.

**Table 7**

*Model Fit Indices for the Measurement Model of AI Literacy*

| Fit Index | χ2(df) | CFI | TLI | RMSEA (90% CI) | SRMR |
|---|---|---|---|---|---|
| Value | 428.390(227)*** | .926 | .918 | .064 (.055-.074) | .054 |

***$p < .001$

The standardized loadings for the first-order factors of AI Literacy ranged from 0.559 to 0.846, with all paths statistically significant except for the first path, which was fixed at 1 (see Table 8).

**Table 8**

*Parameter Estimates for the Measurement Model of AI Literacy*

| | Path | | Estimate | S.E | Beta |
|---|---|---|---|---|---|
| Functional Literacy | → | Q66_1 | 1.000 | 0.000 | 0.672 |
| | → | Q66_2 | 0.923*** | 0.080 | 0.677 |
| | → | Q66_3 | 1.093*** | 0.086 | 0.745 |

| | | | | |
|---|---|---|---|---|
| | → | Q66_4 | 0.917*** | 0.090 | 0.736 |
| | → | Q66_6 | 1.044*** | 0.099 | 0.769 |
| | → | Q66_7 | 1.103*** | 0.102 | 0.802 |
| | → | Q66_11 | 0.617*** | 0.080 | 0.559 |
| | → | Q66_12 | 1.054*** | 0.101 | 0.738 |
| | → | Q66_13 | 0.980*** | 0.097 | 0.709 |
| | → | Q66_14 | 0.859*** | 0.104 | 0.626 |
| Evaluative Literacy | → | Q67_1 | 1.000 | 0.000 | 0.673 |
| | → | Q67_2 | 1.371*** | 0.112 | 0.846 |
| | → | Q67_6 | 1.090*** | 0.103 | 0.690 |
| | → | Q67_7 | 1.458*** | 0.138 | 0.815 |
| | → | Q67_9 | 0.956*** | 0.101 | 0.647 |
| | → | Q67_13 | 1.451*** | 0.157 | 0.785 |
| | → | Q67_14 | 1.489*** | 0.141 | 0.811 |
| Ethical Literacy | → | Q68_1 | 1.000 | 0.000 | 0.673 |
| | → | Q68_3 | 1.170*** | 0.105 | 0.781 |
| | → | Q68_6 | 1.037*** | 0.106 | 0.694 |
| | → | Q68_8 | 0.762*** | 0.120 | 0.607 |
| | → | Q68_9 | 0.707*** | 0.088 | 0.635 |
| | → | Q68_10 | 0.813*** | 0.102 | 0.632 |

***$p < .001$

The estimated covariances among the latent dimensions of FALCON-AI were all statistically significant (see Table 9). While the three dimensions (i.e., functional, evaluative, and ethical) are theoretically distinguishable, their positive associations suggest that they are meaningfully related components of AI Literacy. These findings support the conceptualization of AI Literacy as a multidimensional construct.

**Table 9**

*Correlations Among Latent Variables in the Measurement Model of AI Literacy*

|  | Functional Literacy | Evaluative Literacy |
|---|---|---|
| Evaluative Literacy | 0.876*** |  |
| Ethical Literacy | 0.794*** | 0.783*** |

***p < .001

Next, the final measurement model for Faculty Work is presented in Figure 2.

**Figure 2**

*Measurement Model of Faculty Work*

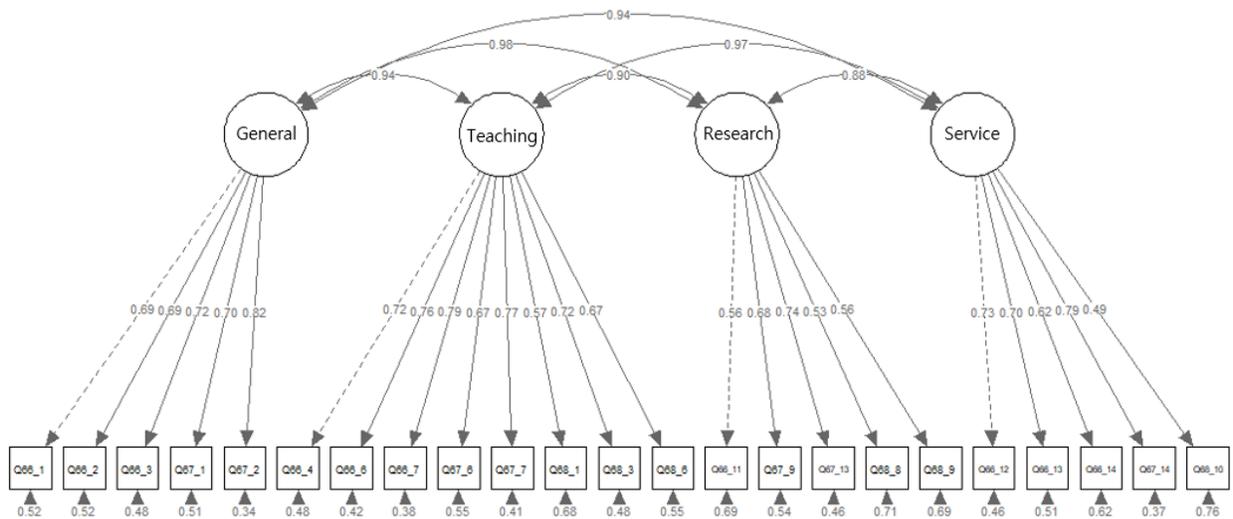

The model fit indices for Faculty Work are shown in Table 10. Although the chi-square test was statistically significant, the TLI and CFI were below 0.90, but the model represents a substantial improvement compared with the original model. The RMSEA exceeded 0.08 but fell below 0.08 within the 90% confidence interval, and the SRMR was below 0.08, suggesting that the model demonstrates a reasonable fit.

**Table 10**

*Model Fit Indices for the Measurement Model of Faculty Work*

| Fit Index | χ2(df) | CFI | TLI | RMSEA (90% CI) | SRMR |
|---|---|---|---|---|---|
| Value | 563.193(224)*** | .876 | .860 | .084 (.076-.093) | .062 |

***$p < .001$

The standardized loadings for the first-order factors of Faculty Work ranged from 0.488 to 0.815, with all paths statistically significant except for the first path, which was fixed at 1 (see Table 11).

**Table 11**

*Parameter Estimates for the Measurement Model of Faculty Work*

| | Path | | Estimate | S.E | Beta |
|---|---|---|---|---|---|
| General | → | Q66_1 | 1.000 | 0.000 | 0.691 |
| | → | Q66_2 | 0.918*** | 0.084 | 0.693 |
| | → | Q66_3 | 1.032*** | 0.08 | 0.724 |
| | → | Q67_1 | 0.785*** | 0.079 | 0.697 |
| | → | Q67_2 | 1.002*** | 0.088 | 0.815 |
| Teaching | → | Q66_4 | 1.000 | 0.000 | 0.722 |
| | → | Q66_6 | 1.155*** | 0.000 | 0.764 |
| | → | Q66_7 | 1.208*** | 0.082 | 0.789 |
| | → | Q67_6 | 0.915*** | 0.096 | 0.668 |
| | → | Q67_7 | 1.195*** | 0.102 | 0.770 |
| | → | Q68_1 | 0.807*** | 0.103 | 0.568 |
| | → | Q68_3 | 1.035*** | 0.098 | 0.723 |
| | → | Q68_6 | 0.96*** | 0.085 | 0.673 |
| Research | → | Q66_11 | 1.000 | 0.000 | 0.558 |
|  | → | Q67_9 | 1.273*** | 0.182 | 0.681 |
|  | → | Q67_13 | 1.721*** | 0.261 | 0.736 |
|  | → | Q68_8 | 0.936*** | 0.175 | 0.535 |
|  | → | Q68_9 | 0.861*** | 0.139 | 0.555 |
| Service | → | Q66_12 | 1.000 | 0.000 | 0.732 |
|  | → | Q66_13 | 0.927*** | 0.072 | 0.701 |
|  | → | Q66_14 | 0.810*** | 0.083 | 0.617 |
|  | → | Q67_14 | 1.087*** | 0.088 | 0.794 |
|  | → | Q68_10 | 0.516*** | 0.075 | 0.488 |

***p < .001

The estimated covariances among the latent dimensions of Faculty Work were all statistically significant (see Table 12). While the four dimensions (e.g., general, teaching, research, and service/administration) are theoretically distinguishable, their positive associations suggest that they are meaningfully related components of Faculty Work. These findings support the conceptualization of Faculty Work as a multidimensional construct.

**Table 12**

*Correlations Among Latent Variables in the Measurement Model of Faculty Work*

|  | General | Teaching | Research |
|---|---|---|---|
| Teaching | 0.937*** |  |  |
| Research | 0.976*** | 0.903*** |  |
| Service | 0.935*** | 0.972*** | 0.885*** |

***p < .001

### 4.3. Reliability

To examine the reliability of the measurement instruments, Cronbach's alpha coefficients were calculated. For AI Literacy, the subscale reliability ranged from 0.83 to 0.90, indicating good internal consistency across all subscales (Tavakol & Dennick, 2011). For Faculty Work, the subscale reliability ranged from 0.75 to 0.89, also demonstrating satisfactory internal consistency. Because the two constructs share the same items, the overall reliability for both AI Literacy and Faculty Work was high, with a Cronbach's alpha of 0.95. These results suggest that the instruments exhibit strong internal consistency at both the subscale and overall scale levels (Table 13).

**Table 13**

*Reliability Coefficients (Cronbach's α) for AI Literacy and Faculty Work*

| Category | Subdimension | Reliability (Cronbach`s alpha) | |
|---|---|---|---|
| AI Literacy | Functional Literacy | 0.91 | 0.95 |
| | Evaluative Literacy | 0.90 | |
| | Ethical Literacy | 0.83 | |
| Faculty work | General | 0.85 | 0.95 |
| | Teaching | 0.89 | |
| | Research | 0.75 | |
| | Service | 0.80 | |

## 5. Conclusion

### 5.1. Summary

This study developed and validated the Faculty Artificial Intelligence Literacy and Competency (FALCON-AI) Scale as a concise, practically ready-to-use, and contextually faculty-centered scale to measure AI literacy in higher education. Grounded in the Critical Technology-resilient Literacy (CTRL) framework, we conceptualized faculty AI literacy as three sub-constructs, namely, functional, evaluative, and ethical literacy. Then, we applied these constructs across four core faculty work domain categories (i.e., general, teaching, research, service/administration), forming a 3 x 4 design framework. The initial pool of 43 items was created by adapting and extending existing AI literacy scales. Then, we conducted expert content validation involving four human subject-matter experts in the related fields and a GPT-based agent as a supplementary reviewer. Revised after the expert content validation process, 40 items were used for the pilot testing, which involved 269 eligible participants from higher education. The results from the pilot testing were subjected to confirmatory factor analysis (CFA) and reliability testing. After prioritizing items with stronger loadings and balancing coverage while reducing respondent burden, the final instrument was reduced to 23 items and demonstrated good model fit for the AI literacy measurement model (CFI = .926, TLI = .918, RMSEA = .064, SRMR = .054) and improved, but comparatively weaker, fit for the faculty work domain model (CFI = .876, TLI = .860, RMSEA = .084, SRMR = .062). Internal consistency was strong, with AI literacy subscales ranging from α = .83–.91, faculty work subscales from α = .75–.89, and overall reliability of α = .95. Collectively, these results support FALCON-AI as a practical and validated tool for diagnosing faculty AI competency in the higher education context.

## 5.2. Contribution

The contributions of this study can be categorized into three domains: theoretical, methodological, and practical. First, theoretically, this study advances the conceptualization of AI literacy in higher education by positioning it as a multidimensional, role-embedded construct grounded in the CRTL framework. Rather than treating AI literacy as a general technical skill or perception-based construct, FALCON-AI adopts an interrelated, comprehensive yet concise CRTL framework including functional, evaluative, and ethical literacy. Moreover, we explicitly situated these three types of literacy within the professional domains of higher education faculty, including teaching, research, and service/administration. This 3 × 4 framework extends prior AI literacy models that primarily target the general public, K-12 or college students, or teachers, and reframes AI literacy as a form of epistemic and professional competency unique to higher education faculty. By empirically supporting the multidimensional structure of AI literacy, this study contributes a theoretically coherent model that bridges technical proficiency, critical judgment, and ethical accountability.

Methodologically, this study demonstrates a rigorous, multi-stage scale development process that integrates theory-driven item generation, expert content validation, GPT-supported triangulation, confirmatory factor analysis (CFA), and reliability testing. An interesting feature of our method was the inclusion of a GPT-based content validation review alongside human subject-matter experts. Quantitatively, inter-rater agreement (IRA) indices showed strong alignment between GPT and human ratings: IRA ranged from .77 to .85 among human experts alone and remained comparably strong (.79 to .87) when GPT ratings were included, indicating that GPT's evaluations of clarity, relevance, and necessity were consistent with expert consensus rather than introducing systematic deviation. Qualitatively, GPT feedback closely mirrored human comments by identifying ambiguous wording, compound items, and misalignment between item phrasing and intended literacy dimensions (e.g., differentiating evaluative from ethical literacy), and by recommending item splitting, clearer operational language, and stronger faculty-specific contextualization. Importantly, GPT did not replace human judgment; instead, it functioned as a triangulation mechanism that enhanced transparency and strengthened confidence in item-level decisions.

Last, this paper contributes to the practical field of AI education by providing a ready-to-use scale to measure AI literacy for higher education faculty (full item list presented in Appendix A). FALCON-AI provides a concise and validated instrument that serves both individual faculty members and higher education institutions. At the individual level, the scale enables faculty to assess their current AI literacy and competency levels, identify strengths and areas for growth across functional, evaluative, and ethical dimensions, and set targeted professional development (PD) goals. By offering a structured reflection tool grounded in authentic faculty responsibilities, FALCON-AI supports intentional skill development rather than ad hoc AI adoption. At the institutional level, the instrument provides data to inform evidence-based decision-making, including identifying institutional competency gaps, prioritizing training areas, and designing faculty PD workshops aligned with actual needs. Furthermore, the scale can be used as a pre- and post-assessment tool within PD workshops to evaluate intervention effectiveness, measure growth over time, and support continuous improvement of AI capacity-building initiatives. In this way, FALCON-AI functions not only as a measurement instrument but also as a strategic tool for guiding responsible and effective AI integration across higher education.

### 5.3. Limitation

Several limitations should be acknowledged. First, the pilot sample, although diverse in roles and disciplines, was recruited through convenience sampling (e.g., social media and professional networks). The convenience-based recruitment strategy may have introduced self-selection bias: participants who chose to participate in the pilot testing may have been more interested in or actively engaged with AI technologies, potentially resulting in higher overall self-reported literacy levels. While such bias does not necessarily compromise factor structure or internal consistency estimates, it may limit the examination of the scale's performance among faculty with minimal AI exposure. Future research should validate the instrument with more systematically sampled populations to examine potential range restriction effects and ensure robustness and generalizability across a varying population of faculty.

Second, the study relied on self-reported Likert-scale responses, which are inherently susceptible to subjective bias, including social desirability effects, overestimation of competence, and variation in how participants interpret AI-related terminology. As a result, FALCON-AI captures perceived AI literacy and competency rather than directly observed performance. While self-assessment is appropriate and common in early-stage scale development, it may not fully reflect faculty members' actual ability to apply AI effectively in authentic professional contexts. Future research could strengthen criterion-related validity by integrating more objective assessment approaches, such as analyzing usage logs or progress data from AI-supported faculty development platforms, incorporating scenario-based or simulation tasks that require faculty to evaluate and revise AI outputs, or conducting structured performance evaluations of AI-mediated teaching, research, or administrative tasks. Combining self-report data with behavioral or task-based measures would enhance the accuracy and objectivity of AI literacy assessment and provide a more comprehensive understanding of faculty competency.

Third, although the AI Literacy measurement model demonstrated good overall fit, the Faculty Work model showed comparatively weaker fit. While the revised model represented improvement over the initial specification and achieved an acceptable level of fit, some factor loadings were relatively low, suggesting that certain items may not have served as particularly strong indicators of their intended latent constructs. However, the items were theoretically derived and content-validated through expert consultation prior to empirical testing, and the CFA was conducted within this predefined framework. Item retention decisions, therefore, reflected a balance between statistical strength and conceptual relevance within the established $3 \times 4$ structure. Future research with larger and more diverse samples is needed to further examine the stability and robustness of the Faculty Work measurement model.

Lastly, a limitation arises due to the rapid evolution of AI technologies and their applications in higher education. Although the FALCON-AI scale was intentionally designed to minimize obsolescence by avoiding references to specific AI tools or platforms and using transferable competencies (e.g., functional use, evaluative judgment, and ethical decision-making), the ways in which AI is integrated into teaching, research, and administrative practice are likely to change quickly. As new capabilities emerge and institutional policies evolve, certain operational practices or expectations embedded in the items may require revision to remain up-to-date and contextually relevant. Therefore, FALCON-AI should be viewed as a living instrument rather than a static tool. We plan to periodically revisit the instrument and update the scale to ensure alignment with technological developments, emerging ethical considerations, and shifting professional standards in higher education.

# Appendix

*FALCON-AI Scale Questionnaire*

Please indicate how much you agree with each statement below. (23 questions in total.)
1: Strongly disagree
2: Somewhat disagree

3: Neither agree nor disagree
4: Somewhat agree
5: Strongly agree

| Domain | Number | Item |
|---|---|---|
| Functional Literacy | FL-1 | I can explain core concepts of AI (e.g., machine learning, generative models). |
| | FL-2 | I can design and refine prompts to guide AI toward generating useful content. |
| | FL-3 | I can customize an AI agent for specific purposes. |
| | FL-4 | I can use AI to generate teaching materials. |
| | FL-5 | I can use AI to support accessibility of instruction (e.g., real-time transcription). |
| | FL-6 | I can use AI to support personalization of instruction (e.g., differentiation). |
| | FL-7 | I can use AI as a writing assistant. |
| | FL-8 | I can use AI to automate repetitive administrative tasks (e.g., answering FAQs). |
| | FL-9 | I can use AI to track service activities (e.g., writing reports, meeting notes). |
| | FL-10 | I can use AI to assist in drafting language for regulatory documents (e.g., IRB). |
| Evaluative Literacy | Ev-1 | I know strengths and weaknesses of common AI tools. |
| | Ev-2 | I can choose appropriate AI tools for particular tasks. |
| | Ev-3 | I can critically assess the pedagogical contribution of AI to my teaching field. |
| | Ev-4 | I can compare different AI tools and determine which ones to use for different teaching tasks (e.g., lesson design, student feedback). |
| | Ev-5 | I can critically evaluate the implications of AI applications in research. |
| | Ev-6 | I can compare different AI tools and determine which ones to use for different research tasks (e.g., brainstorming, literature review, data analysis, writing assistance) |
| | Ev-7 | I use different AI tools for different administrative work, considering their strengths. |
| Ethical Literacy | Eth-1 | I can include a clear AI guideline in my syllabus detailing how students may or may not use AI for assignments and assessments. |
| | Eth-2 | I can design instructional activities so that students can consider how they should or should not integrate AI |
| | Eth-3 | I can design assessments that remain valid and fair in environments where AI tools are widely available. |
| | Eth-4 | I am aware of the risks of uploading participant data to AI for analysis. |
| | Eth-5 | I know the need for properly attributing the use of AI in scholarly writing and data analysis. |
| | Eth-6 | I can identify the ethical concerns regarding the deployment of AI at the departmental or college level. |